\documentclass{PoS}

\usepackage{bbm,amsmath} 

\newcommand{\fm}{\, {\rm fm}}

\newcommand{\gev}{\, {\rm GeV}}
\newcommand{\mev}{\, {\rm MeV}}

\newcommand{\MSbar}{\overline{\rm MS}}

\newcommand{\be}{\begin{equation}}
\newcommand{\ee}{\end{equation}}
\newcommand{\bea}{\begin{eqnarray}}
\newcommand{\eea}{\end{eqnarray}}

\newcommand{\bi}{\begin{itemize}}
\newcommand{\ei}{\end{itemize}}

\boldmath
\title{B-physics from the ratio method with Wilson twisted mass fermions}

\unboldmath

\ShortTitle{B-physics from the ratio method}
\author{
\vspace*{-5mm}
\begin{flushright}
IFIC/12-71\\
FTUAM-12-107\\
IFT-UAM/CSIC-12-X1\\
RM3-TH/12-16\\
LTH960\\
HU-EP-12/37
\end{flushright}
N. Carrasco$^{(a)}$, P. Dimopoulos$^{(b,c)}$, R. Frezzotti$^{(b,c)}$, V. Gim\'enez$^{(a)}$, 
G. Herdoiza$^{(d)}$, \newline 
V. Lubicz$^{(e,f)}$, G. Martinelli$^{(g,h)}$, C. Michael$^{(i)}$, 
D. Palao$^{(c)}$, G.C. Rossi$^{(b,c)}$,\newline 
F. Sanfilippo$^{(j)}$, \speaker{A.~Shindler}\thanks{Heisenberg Fellow}$~~^{(k)}$, 
S. Simula$^{(f)}$, C. Tarantino$^{(e,f)}$
\\
$^{(a)}$Departament de F\'isica Te\`orica and IFIC, Univ. de Val\`encia-CSIC 
\vspace{0.2cm}
\\
$^{(b)}$Dipartimento di Fisica, Universit\`a 
di Roma ``Tor Vergata'' 
\vspace{0.2cm}
\\
$^{(c)}$INFN, Sezione di ``Tor Vergata'' c/o Dipartimento di Fisica, 
Universita` di Roma ``Tor Vergata'' 
\vspace{0.2cm}
\\
$^{(d)}$Departamento de F\'isica Te\`orica and Instituto de F\'isica Te\`orica UAM/CSIC, 
\vspace{0.2cm}
\\
$^{(e)}$Dipartimento di Fisica, Universit\`a Roma Tre 
\vspace{0.2cm}
\\
$^{(f)}$INFN, Sezione di Roma Tre c/o Dipartimento di Fisica, Universit\`a Roma Tre 
\vspace{0.2cm}
\\
$^{(g)}$SISSA 
\vspace{0.2cm}
\\
$^{(h)}$INFN, Sezione di Roma, 
\vspace{0.2cm}
\\
$^{(i)}$Theoretical Physics Division, Dept. of Mathematical Sciences, 
University of Liverpool
\vspace{0.2cm}
\\
$^{(j)}$Laboratoire de Physique Th\'eorique (Bat. 210), Universit\'e Paris Sud, 
\vspace{0.2cm}
\\
$^{(k)}$Institut f\"ur Elementarteilchenphysik, Fachbereich Physik, 
Humbolt Universit\"at zu Berlin, 
}

\abstract{We present a precise lattice QCD determination of the $b$-quark mass, of the $B$ and $B_s$ decay constants
and first preliminary results for the $B$-mesons bag parameter. 
Simulations are performed with $N_f$ = 2 Wilson twisted mass fermions 
at four values of the lattice spacing and the results are extrapolated to the continuum limit.
Our calculation benefits from the use of improved interpolating operators 
for the $B$-mesons and employs the so-called ratio method.
The latter allows a controlled interpolation at the $b$-quark mass between 
the relativistic data around and above the charm quark mass and the exactly known static limit.
}

\FullConference{The 30 International Symposium on Lattice Field Theory - Lattice 2012,\\
June 24-29, 2012\\
Cairns, Australia}

\begin{document}
\section{Introduction}
\vspace{-0.4cm}
Stringent tests of the Standard Model and the search for new physics 
pass through a detailed study of physical processes involving the $b$-quark. 
It is clear that, despite the present (B-factories, LHCb) and future (SuperB factories) experimental programs,
lattice QCD results of hadronic parameters need to have reduced uncertainties at the level of $\sim 1 \%$
(see refs~\cite{Tarantino:2012mq,Zanotti:2012ic} for recent reviews).
In this proceeding we report on the ongoing project, within the ETMC collaboration, to compute
B-physics hadronic parameters with Wilson twisted mass fermions.
We extend and improve our previous 
analysis~\cite{Dimopoulos:2011gx,Blossier:2009hg} in two ways:
\begin{itemize}
\vspace{-0.2cm}
\item we optimize the interpolating operators for heavy-light systems to better project onto the
fundamental state
\vspace{-0.2cm}
\item we extend the range of heavy masses, $\mu_h$, considered reaching values of approximately 
$\mu_h \sim 2.5\,\mu_c$, where $\mu_c$ is the charm quark mass.
\end{itemize}
\vspace{-0.6cm}
\section{Improved interpolating operators}
\label{sec:impr_op}
\vspace{-0.4cm}
We use, for this analysis, the $N_f=2$ dynamical gauge configurations with up and down mass degenerate 
quarks, $\mu_{u/d}=\mu_l$, generated by the European 
Twisted Mass Collaboration (ETMC)~\cite{Baron:2009wt}. 
The lattice action is the tree-level improved Symanzik gauge action~\cite{Weisz:1982zw} 
and the twisted mass quark action~\cite{Frezzotti:2000nk} at maximal twist~\cite{Frezzotti:2003ni}.
The strange, $\mu_s$, and the charm (heavy), $\mu_c$ ($\mu_h$), quarks are quenched in this work. \newline
We have used four lattice spacings $a=\{0.098(3), 0.085(2), 0.067(2), 0.054(1)\}\,\fm$~\cite{Blossier:2010cr} 
corresponding to~$\beta=\{3.80, 3.90, 4.05, 4.20\}$. We also use the values of the renormalized
quark masses $\overline\mu_{u/d}=3.6(2)\,\mev$ and $\overline\mu_s=95(6)\,\mev$~\cite{Blossier:2010cr} 
and the pseudoscalar density renormalization constants
$Z_P^{\MSbar}(2\gev)=\{0.411(12), 0.437(7), 0.477(6), 0.501(20)\}$~\cite{Constantinou:2010gr} at the four beta values.
We denote by a ``bar'' the quark masses renormalized in the $\overline{MS}$ scheme 
at a renormalization scale of $2$ GeV.
The values of the bare valence quark masses used in this calculation are given in tab.~\ref{tab:sim_points}.

To keep the noise-to-signal ratio under control we extract meson masses
at relatively small temporal separations. To improve the projection onto the 
fundamental state, we need trial states with large overlap with the lowest lying energy eigenstate.
We have constructed a smeared interpolating operator using Gaussian smearing~\cite{Gusken:1989qx}, i.e.
we have inverted on a source $\Phi^{\rm S}$
\be
\Phi^{\rm S} \propto \left(1+\kappa_{\rm G}a^2\nabla^2_{\rm APE}\right)^{N_{\rm G}}\Phi^{\rm L}\,, \qquad \kappa_{\rm G} = 4 
\quad N_{\rm G} = 30\,,
\ee
where $\Phi^{\rm L}$ is the standard local source and $\nabla_{\rm APE}$ is the lattice covariant derivative
with APE smeared gauge links ($\alpha_{\rm APE}=0.5$, $N_{\rm APE}=20$)~\cite{Jansen:2008si}.
We have used both smeared sources and sinks finding that the best correlator in respect of signal to noise
ratio is the smeared-local (SL) correlator where the source is smeared and the sink is local.

\begin{figure}[t]
\vspace{-0.9cm}
\hspace{0.3cm}\includegraphics[width=0.5\textwidth,angle=0]{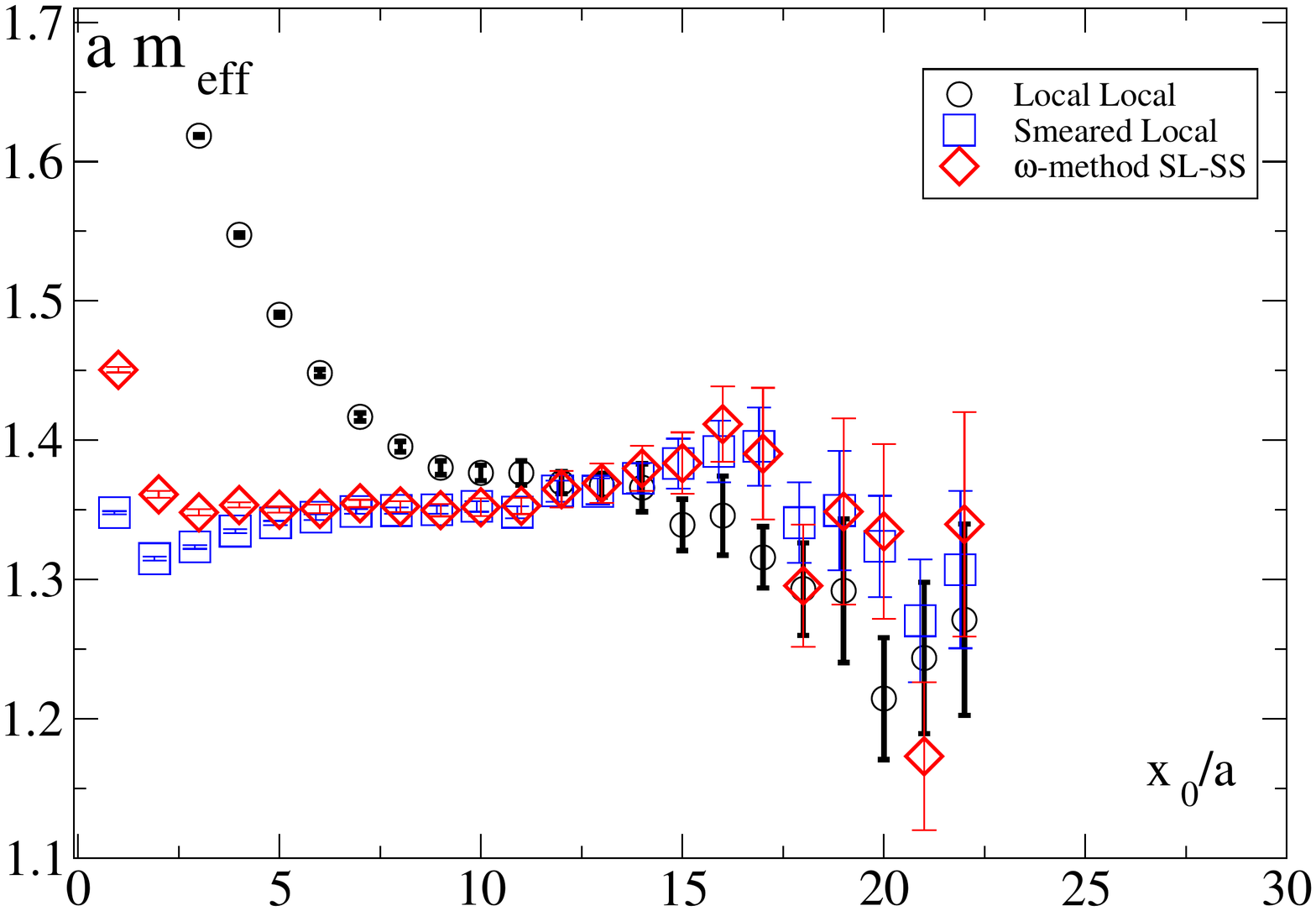}
\vspace{-0.5cm}
\includegraphics[width=0.5\textwidth,angle=0]{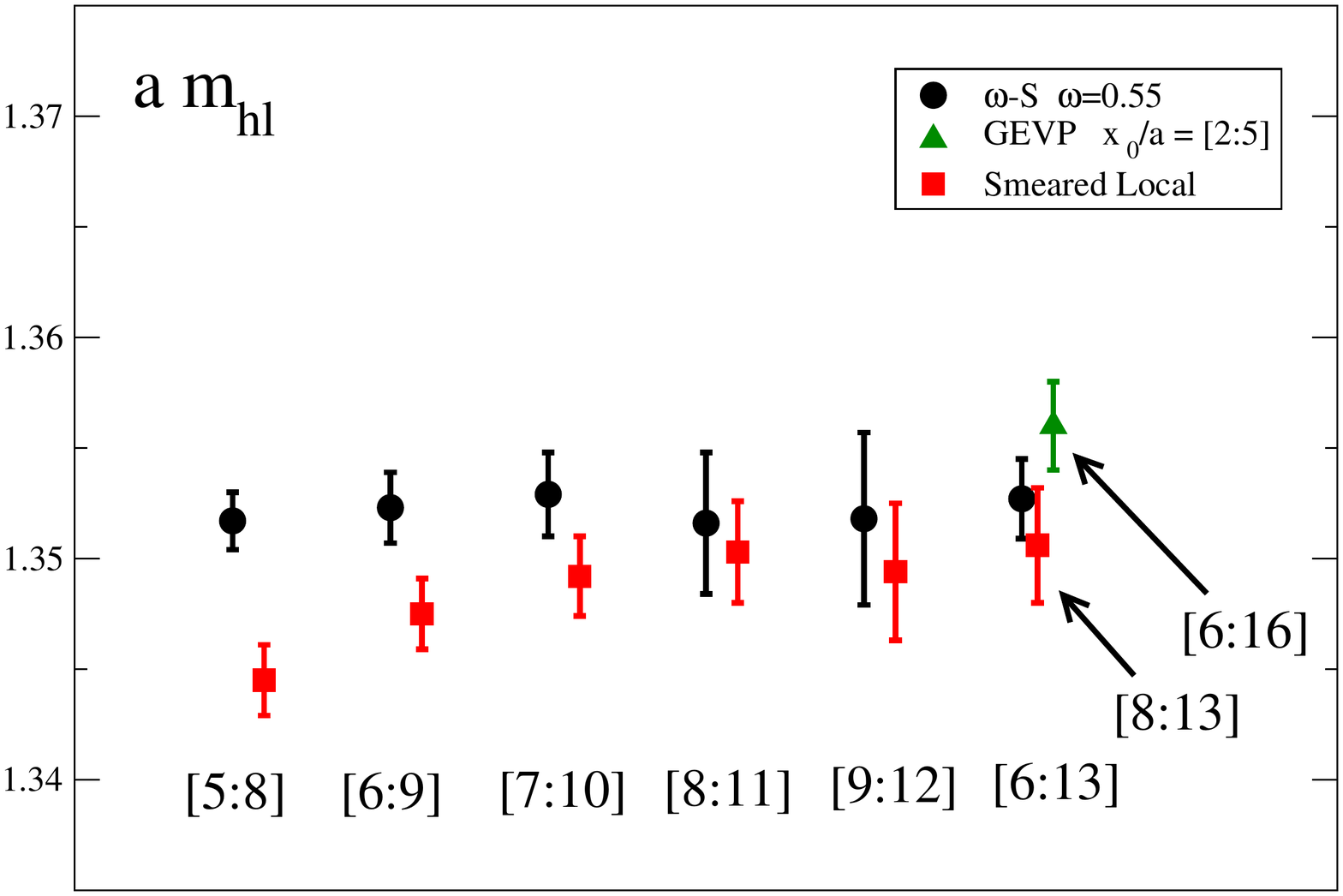}
\caption{Left plot: Euclidean time dependence of the 
effective masses at $\beta=3.8$, $a\mu_l=0.08$, $a\mu_h=0.5246$
obtained from LL, SL and $\omega$S correlation functions. Right plot: 
effective masses at the same simulation point as for the left plot
extracted from the SL and the $\omega$S correlation for different plateau ranges. 
The ranges, in units of the lattice spacing $a$, are indicated in square brackets. We also plot for comparison
the result of a GEVP analysis.}
\label{fig:eff_masses}
\end{figure}

To further improve the overlap with the ground state of the B-system we have constructed
the source
\be
\Phi(\omega) \propto \omega \Phi^{\rm S} + (1-\omega)\Phi^{\rm L}\,,
\ee
dependent on a tunable parameter $\omega$. We have computed correlators where the source is
$\Phi(\omega)$ and the sink is smeared ($\omega$S) for several values of $\omega$. We then optimize
the value of $\omega$ to achieve a projection onto the ground state at earlier Euclidean times than
with SL correlators
\footnote{Details on the optimization procedure will be given in a forthcoming publication.}.
\begin{table}[!t]
\begin{tabular}{||c||c|c|c||}
\hline
$\beta$  &  $a \mu_\ell$ &   $a \mu_s$    & $a \mu_h$ \\ \hline\hline
3.80 & 0.0080, 0.0110 & 0.0175, 0.0194 & 0.1982, 0.2331, 0.2742, 0.3225, 0.3793 
                                                0.4461 \\ 
     &                &     0.0213     &  0.5246 \\ \hline
3.90 &     0.0030, 0.0040     &  0.0159, 0.0177 & 0.1828, 0.2150, 0.2529, 0.2974, 0.3498 \\
     & 0.0064, 0.0085, 0.0100 &      0.0195     & 0.4114, 0.4839  \\ \hline
4.05 & 0.0030, 0.0060  & 0.0139, 0.0154 & 0.1572, 0.1849, 0.2175, 0.2558, 0.3008 \\
     &     0.0080      &     0.0169     & 0.3538, 0.4162 \\ \hline
4.20 & 0.0020, 0.0065 & 0.0116, 0.0129   & 0.13315, 0.1566, ,0.1842, 0.2166, 0.2548 \\
     &                &     0.0142       & 0.2997, 0.3525 \\ \hline
\hline
\end{tabular}
\caption{Bare parameters used in this computation. The volumes $L^3 \times T$ simulated are $L/a=\{24, 24, 32, 48\}$, with $T/L=2$,
respectively for $\beta=\{3.80, 3.90, 4.05, 4.20\}$. Only for the ensemble with $a\mu_\ell = 0.003$ and $\beta=3.9$ we have $L/a=32$.}
\label{tab:sim_points}
\end{table}
In the left plot of fig.~\ref{fig:eff_masses} we show the Euclidean time dependence of the 
effective masses for a particular simulation
point obtained from local-local (LL), SL and $\omega$S correlation functions.
The plot describes well the general properties for heavy-light effective masses: the SL correlators
are already substantially improved for a safe extraction of the ground state mass compared with the 
LL correlators. The $\omega$S correlation functions perform even better.
This can be fully appreciated in the right plot of fig.~\ref{fig:eff_masses} where we compare for different
plateau regions the effective masses extracted from the SL and the $\omega$S correlation
functions. It is clear that with $\omega$S correlation functions we can extract the mass 
of the ground state using results at Euclidean times smaller than with the SL ones.
In the same plot we also compare these two determinations with the one obtained
with the GEVP~\cite{Blossier:2009kd}, obtaining perfectly consistent results within statistical uncertainties.
We prefer not to use the GEVP in our final analysis because we have found that it gives noisier
estimates for the pseudoscalar decay constant. All the results presented in the 
following sections take advantage of these improvements.
\vspace{-0.6cm}
\section{The $b$-quark mass and decay constants $f_B$ and $f_{B_s}$}
\begin{figure}[tb]
\vspace{-0.9cm}
\hspace{0.3cm}\includegraphics[width=0.5\textwidth,angle=0]{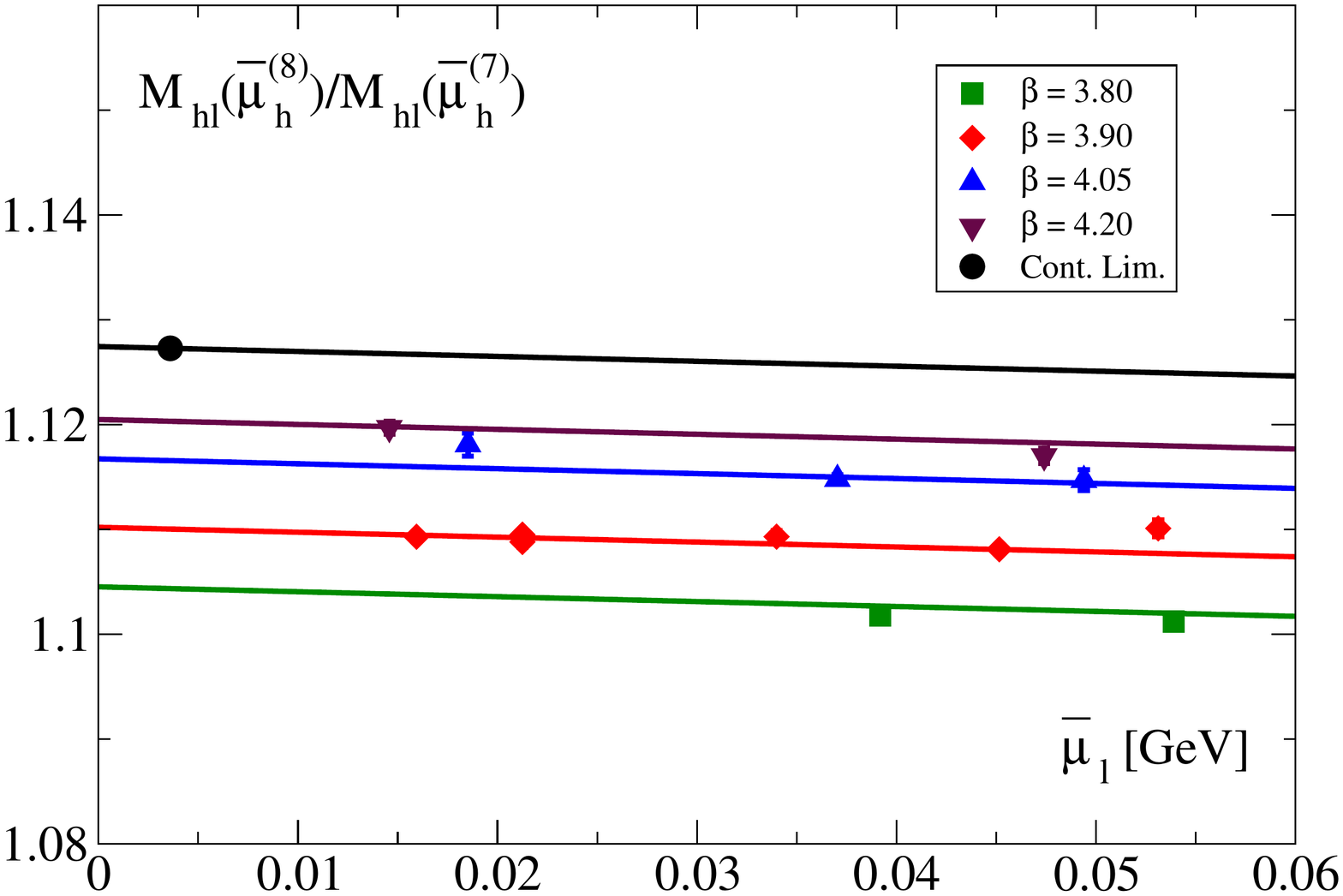}
\vspace{-0.5cm}
\includegraphics[width=0.5\textwidth,angle=0]{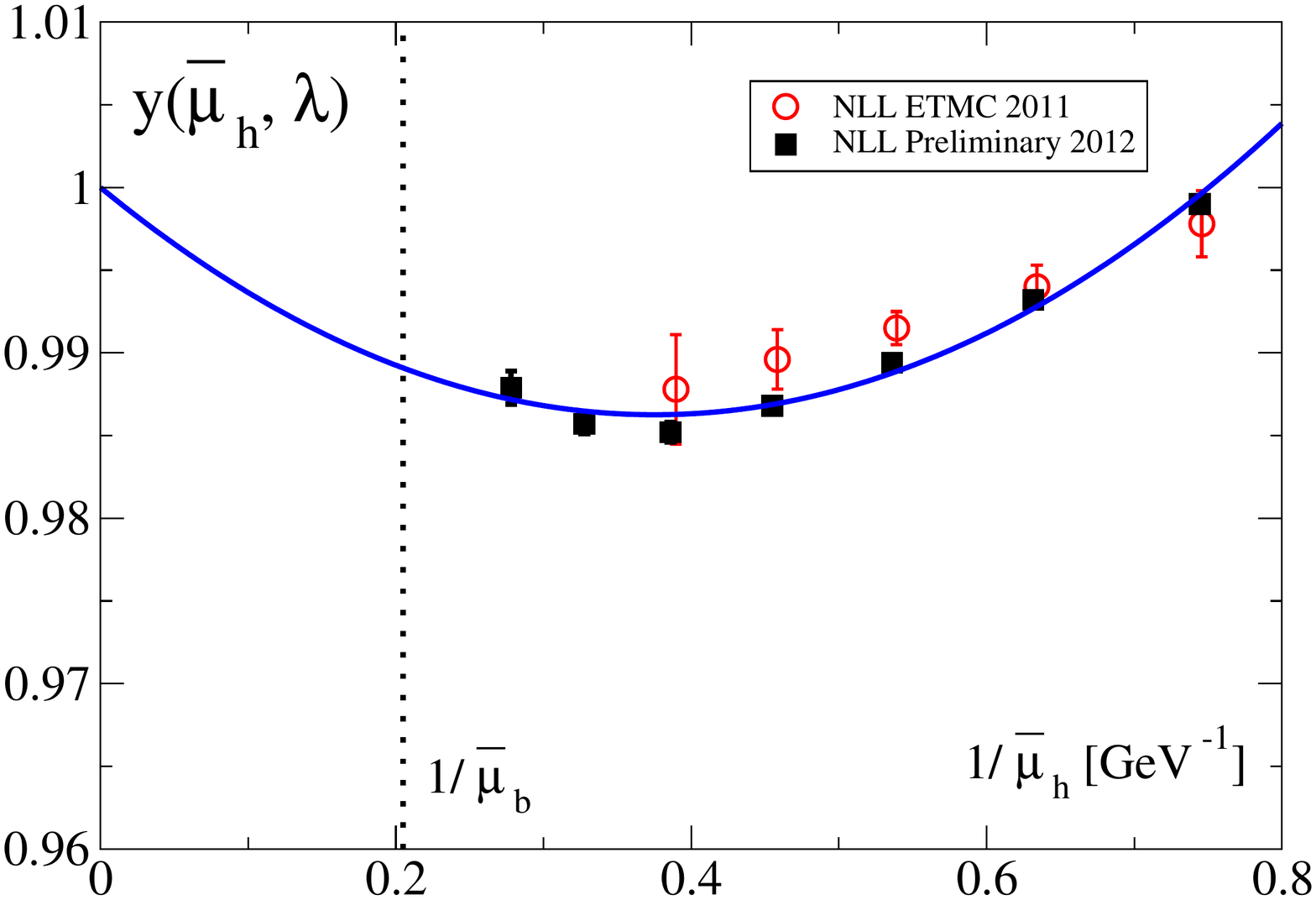}
\caption{Left plot: chiral-continuum extrapolation of the ratio 
of heavy-light meson masses at 
the heaviest quark masses, $\overline\mu_h^{(7,8)}$. Right plot: non-perturbative
heavy-quark mass dependence of the ratio function $y(\overline\mu_h,\lambda)$, for $\lambda= 1.1784$ in the continuum. 
For comparison we show our previous results~\cite{Dimopoulos:2011gx} without the interpolating operators improvement discussed
in sect.~{\protect\ref{sec:impr_op}}.}
\label{fig:b_mass}
\end{figure}

\vspace{-0.4cm}
To determine the $b$-quark mass, $f_B$, $f_{B_s}$ and the bag parameters, we implement the ratio method~\cite{Blossier:2009hg}. 
We refer to this ref. for the details of the method. 
Here we briefly recall the basic steps
for the determination of the $b$-quark mass. For the other quantities discussed in these
proceedings the same strategy applies.
The method is suggested by the HQET asymptotic behavior of the heavy-light
meson mass $M_{hl}$ in the pole heavy-quark mass $\mu_h^{\rm pole}$
\be
\left(\frac{M_{hl}}{\mu_h^{\rm pole}}\right) = 1 + O\left(\frac{1}{\left(\mu_h^{\rm pole}\right)} \right)\,.
\label{eq:Mhl}
\ee
We consider a sequence of heavy-quark masses $\overline\mu_h^{(1)}<\overline\mu_h^{(2)} < \cdots<\overline\mu_h^{(N)}$
with fixed ratio $\lambda$, i.e. $\overline\mu_h^{(n)} = \lambda \overline\mu_h^{(n-1)}$.
The key quantities of the method are ratios of heavy-light
meson masses at subsequent values of the heavy-quark mass, $y(\overline\mu_h^{(n)},\lambda;\overline\mu_l,a)$, 
properly normalized~\cite{Blossier:2009hg}.
\label{sec:bmass}
From eq.~\eqref{eq:Mhl} and QCD asymptotic freedom it follows that the ratios 
$y(\overline\mu_h^{(n)},\lambda;\overline\mu_l,a)$ have an exact static limit:
\be
\lim_{\overline\mu_h \rightarrow \infty} \lim_{a\rightarrow 0}y(\overline\mu_h^{(n)},\lambda;\overline\mu_l,a) = 1\,.
\ee
The value of $\lambda$ is chosen in such a way that after a finite number of steps 
the heavy-light meson mass 
assumes the experimental value $M_B = 5.279$ GeV.
In order to implement this condition, the lattice data at the four lattice spacings 
are interpolated at specific values of the heavy-quark mass.
In the left plot of fig.~\ref{fig:b_mass} we show the chiral and continuum
extrapolation of the ratio of the heavy-light meson masses evaluated at the heaviest 
quark mass.
A phenomenological fit linear in the light-quark mass and in $a^2$
described the lattice data rather well.
We observe that discretization errors are well under control.
The values of the ratios $y(\overline\mu_h,\lambda)$ extrapolated to the chiral 
and continuum limit have a non-perturbative heavy-quark mass dependence that is well described 
by the HQET-inspired function
\be
y(\overline\mu_h,\lambda) = 1+\frac{\eta_1(\lambda)}{\overline\mu_h} + \frac{\eta_2(\lambda)}{\overline\mu_h^2}\,,
\label{eq:fit_ansatz}
\ee
as it can be seen from the right plot in fig.~\ref{fig:b_mass}.
From this plot it is clear that the improvements we have implemented in this analysis
allow to have an accurate description of the non-perturbative dependence on $\overline\mu_h$
of the ratio $y$. In particular there is little doubt that the numerical data are well
described by the fit ansatz in eq.~\eqref{eq:fit_ansatz}.
The formula~\eqref{eq:fit_ansatz}, with the fit parameters $\eta_1$ and $\eta_2$, provides 
the non-perturbative description of the heavy-quark mass dependence of the ratio $y(\overline\mu_h,\lambda)$
over the whole range of heavy-masses.
We can thus compute the value of the $b$-quark mass simply using the knowledge of $y(\overline\mu_h,\lambda)$
to iterate the value of the triggering heavy-light meson mass $M_{hl}(\overline\mu_h^{(1)})$ from the charm region
to the experimental value of the $B$ meson mass $M_B$.
This is always possible
with a slight tuning of $\lambda$ and $\overline\mu_h^{(1)}$. One finds
$
\lambda = 1.1784, \, \overline\mu_h^{(1)} = 1.14 {\rm~GeV} 
\Rightarrow \overline\mu_b = \lambda^K \overline\mu_h^{(1)} (K=9)\,,
$
that leads after a 4-loop evolution to
\be
\mu_b^{\overline{MS}}(\mu_b)|_{N_f=2} = 4.35(12) {\rm~GeV}\,.
\ee
This result is consistent with our previous determination~\cite{Dimopoulos:2011gx} with
slightly reduced error, and a fully equivalent one is obtained if we use
as input the heavy-strange meson mass $M_{hs}$.
The $2.7\%$ relative error is dominated by the ones associated to the scale
and $Z_P$ determinations. Note that a $b$-quark mass value smaller by $\sim
120$~MeV would be found by setting $N_f=4$ (rather than $2$) in the
evolution from $2$ GeV to the $b$-scale.
The statistical and systematic uncertainties coming from the application of the ratio method
turn out to be negligible, after the improvements just discussed.

\begin{figure}[tb]
\vspace{-0.9cm}
\hspace{0.3cm}\includegraphics[width=0.5\textwidth,angle=0]{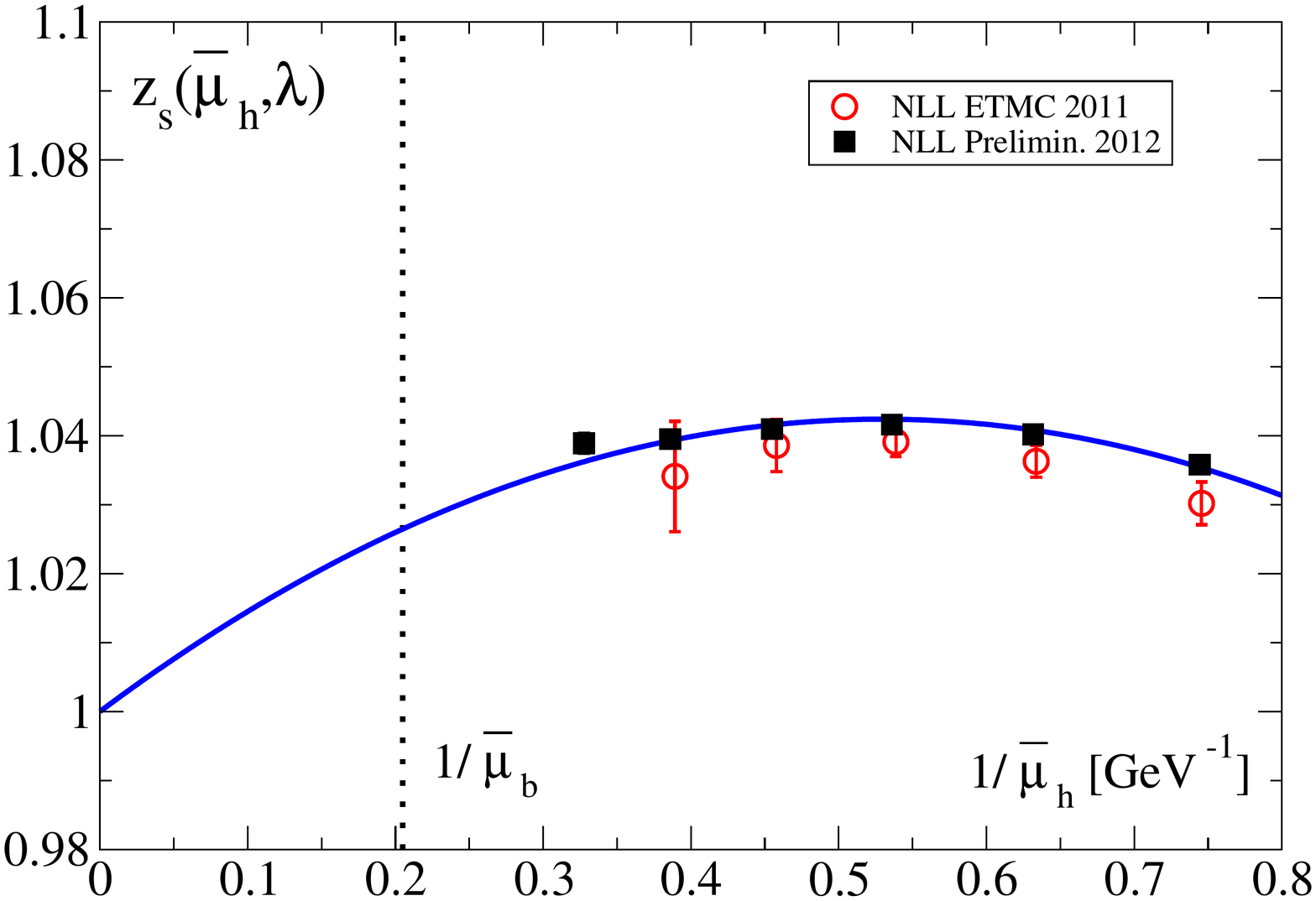}
\vspace{-0.5cm}
\includegraphics[width=0.5\textwidth,angle=0]{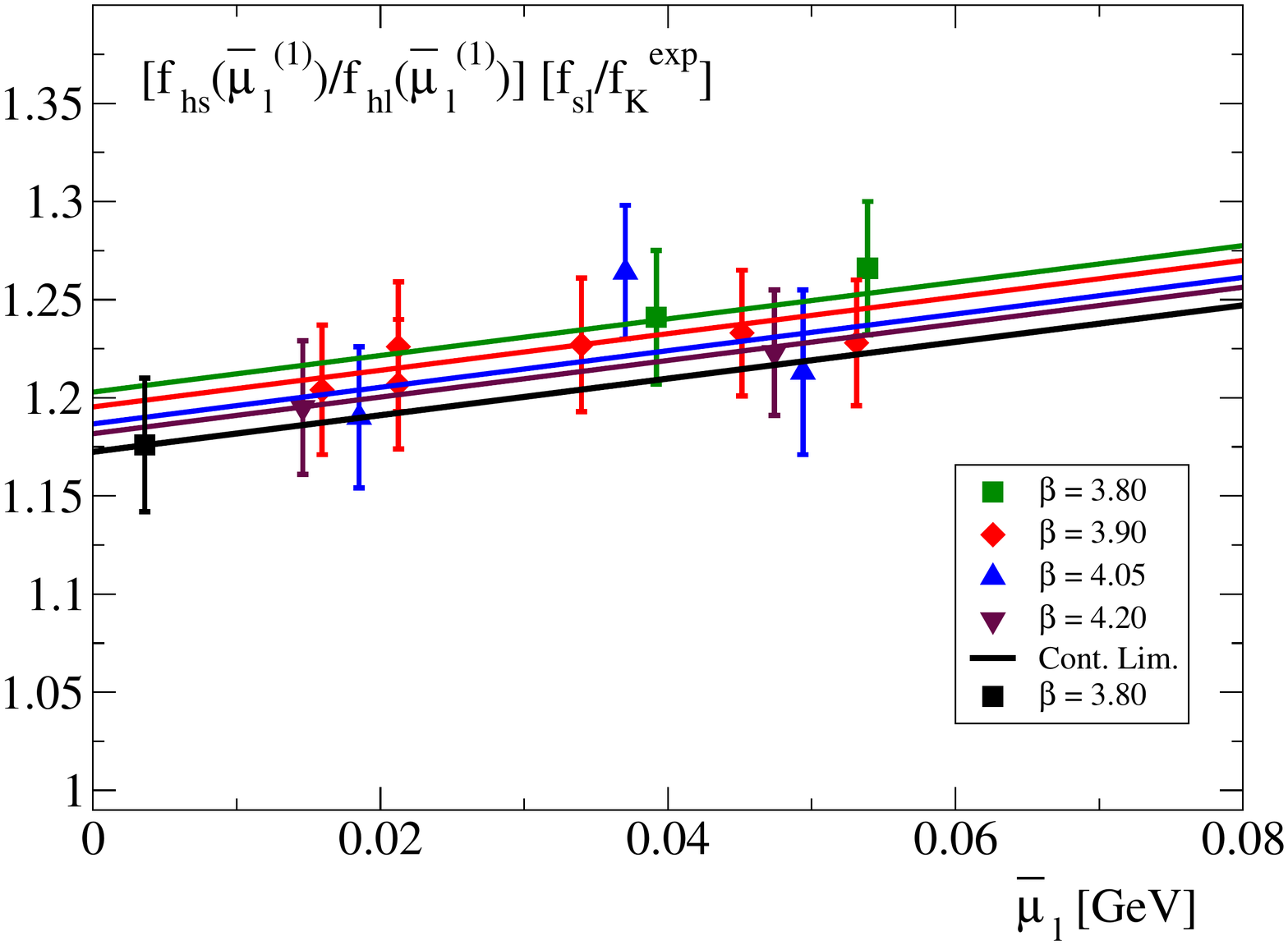}
\caption{Left plot: same as the right plot in fig.~{\protect\ref{fig:b_mass}} fot the ratio function $z_s(\overline\mu_h,\lambda)$.
Right plot: chiral extrapolation for the double ratio $\left[f_{hs}/f_{hl}\right]\cdot\left[f_{sl}/f_K^{\rm exp} \right]$ 
at the triggering point, $\overline\mu_h^{(1)}$.}
\label{fig:decay_constant}
\end{figure}

To compute heavy-light and heavy-strange decay constants a completely analogous strategy can be adopted.
We find advantageous to define ratios with exactly known static limit for the heavy-strange decay constant,
$z_s(\overline\mu_h^{(n)},\lambda;\overline\mu_l,\overline\mu_s,a)$, and for the ratio $f_{B_s}/f_B$, 
$\zeta(\overline\mu_h^{(n)},\lambda;\overline\mu_l,\overline\mu_s,a)$~\cite{Dimopoulos:2011gx}.
As for the $b$-quark mass, $f_{hs}$ at the triggering point 
and the ratio $z_s$ have a smooth chiral-continuum extrapolation
with cutoff effects always well under control. The heavy-quark mass dependence in the continuum
of $z_s$ can again be described by a formula as the one in eq.~\eqref{eq:fit_ansatz},
as shown in the left plot of fig.~\ref{fig:decay_constant}. 
The application of the ratio method using as input the
value we determined for $\overline\mu_b$ leads us to 
\be
f_{B_s} = 234(6) {\rm ~MeV}\,.
\ee

The chiral-continuum extrapolation for the double ratio 
$\zeta(\overline\mu_h^{(n)},\lambda;\overline\mu_l,\overline\mu_s,a)$ does not pose any problem and its
heavy-quark mass dependence is very weak easing the interpolation at the $b$-quark mass. 
At the triggering mass, $\bar{\mu}_h^{(1)}$, the data for $f_{hs}/f_{hl}$ as a function of $\bar\mu_l$ suggest
that the $h$-quark still behaves as a relativistic one. Thus for the chiral-continuum extrapolation
of $f_{hs}/f_{hl}(\bar{\mu}_h^{(1)})$ we find convenient to study the double ratio
$\left[ f_{hs} / f_{hl} \right] \cdot \left[f_{sl} / f_K^{\rm exp}\right]$, for which a smooth
dependence on the light-quark mass is expected~\cite{Roessl:1999iu,Allton:2008pn}.
We find that a linear fit in the light-quark mass describes well the data as it can be seen from the
right plot in fig.~\ref{fig:decay_constant}.
This leads to the values
\be
\frac{f_{B_s}}{f_B} = 1.19(5), \qquad f_B = 197(10) {\rm~MeV}\,,
\ee
where we add in quadrature the statistical and systemtic uncertainties.
\vspace{-0.5cm}
\section{Bag parameters}
\label{sec:bag}
\begin{figure}[tb]
\vspace{-0.8cm}
\hspace{0.3cm}\includegraphics[width=0.5\textwidth,angle=0]{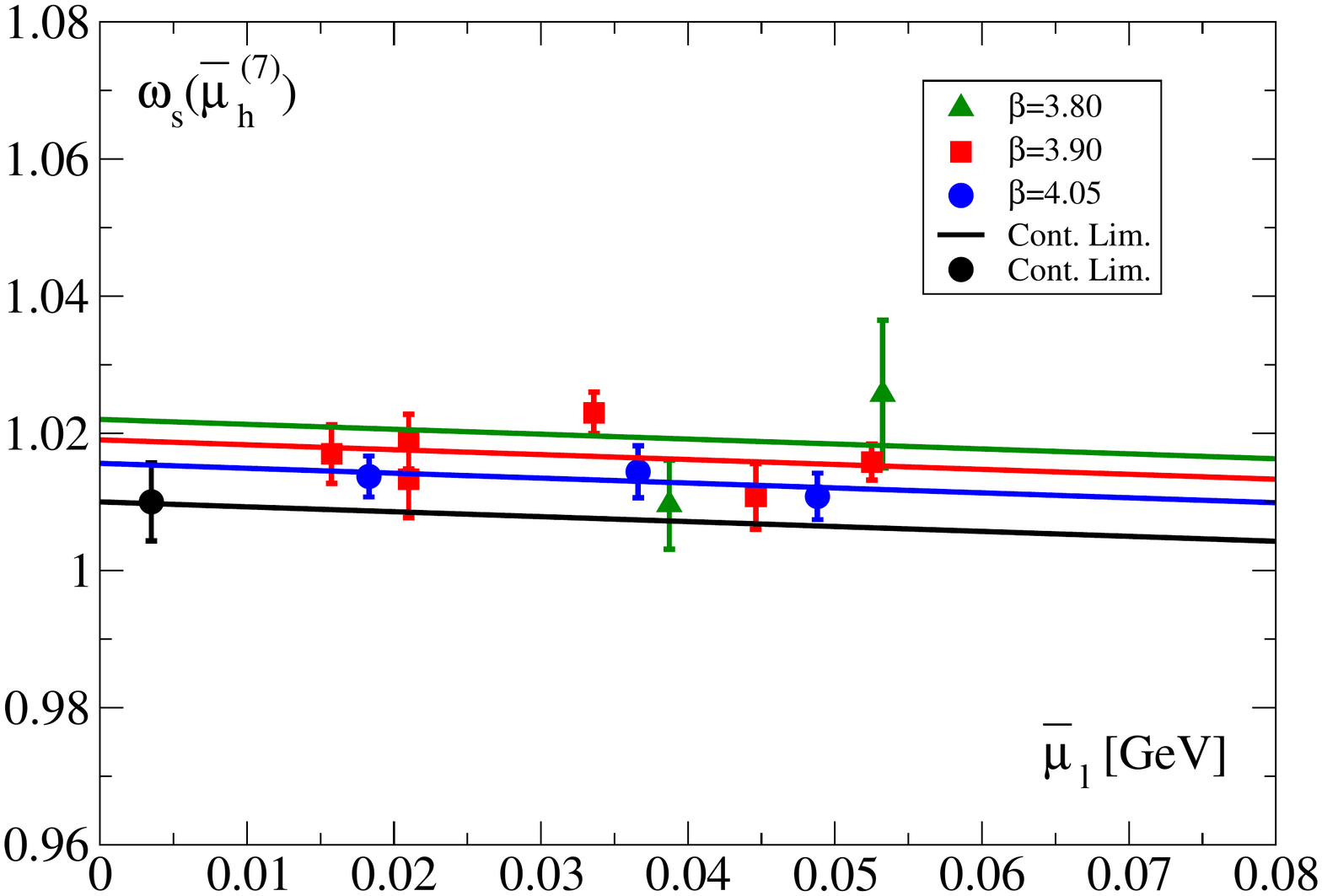}
\vspace{-0.5cm}
\includegraphics[width=0.5\textwidth,angle=0]{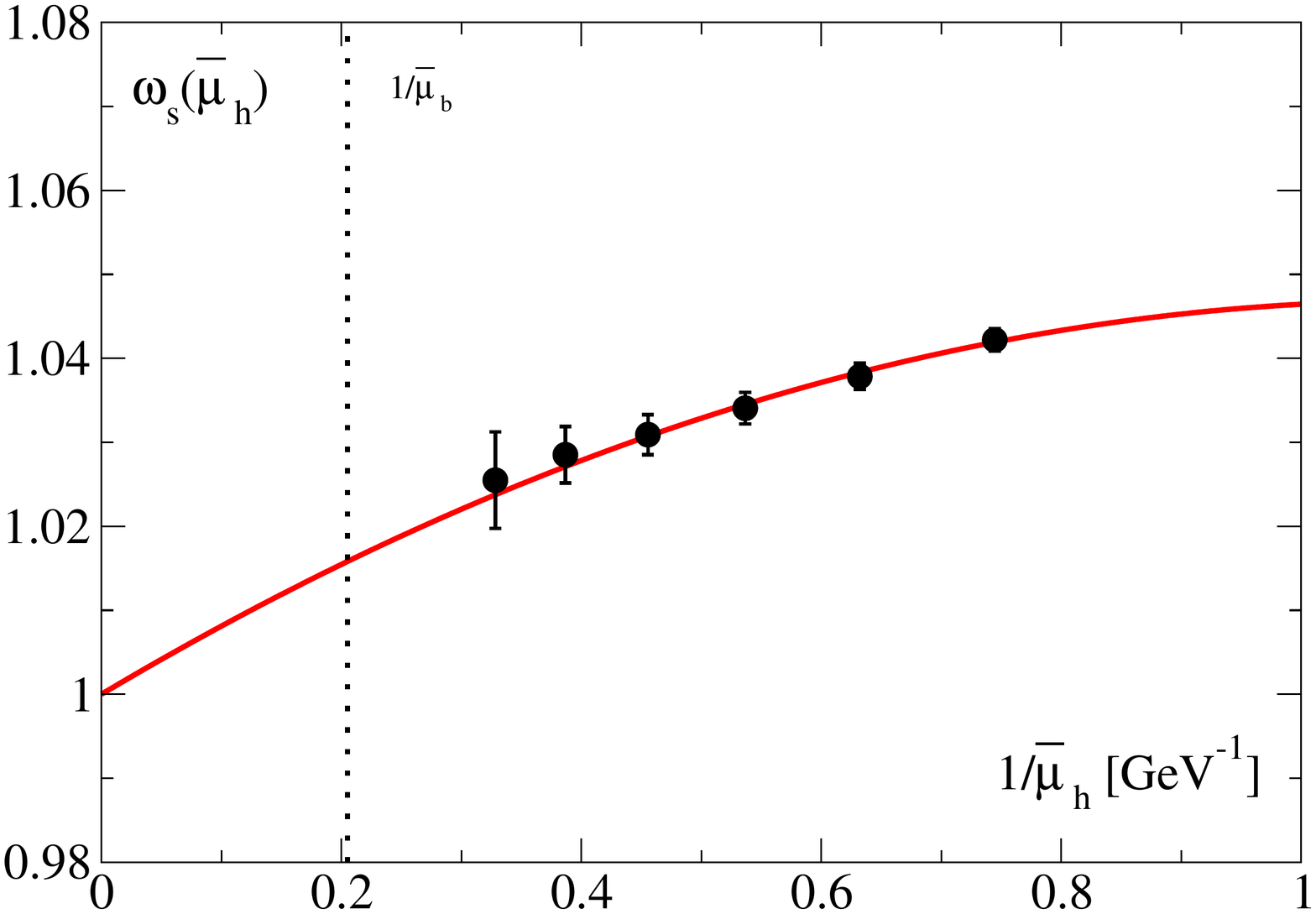}
\caption{Left plot: chiral-continuum extrapolation of the ratio 
of the bag parameters, $\omega_s$, for
the heaviest quark mass analyzed, $\overline\mu_h^{(7)}$. Right plot: non-perturbative
heavy-quark mass dependence of the ratio function $\omega_s(\overline\mu_h,\lambda)$, for $\lambda= 1.1784$ in the continuum.}
\label{fig:bag}
\end{figure}

\vspace{-0.4cm}
The $B$-parameter of the renormalized operator in QCD is defined as 
\be
\langle \overline{B}_q | O^{\Delta B=2}_q | B_q \rangle^{\overline{MS}} \equiv \frac{8}{3}f_{B_q}^2 B_{B_q}^2(\mu) M_{B_q}^2
\ee
where $O^{\Delta B=2}_q \equiv \left[\overline{b}\gamma_\mu (1 - \gamma_5) q\right]\left[\overline{b}\gamma_\mu (1 - \gamma_5) q \right]$ and $\mu$ is the renormalization scale.
With Wtm fermions the renormalization of this operator is multiplicative~\cite{Frezzotti:2004wz}.
The ratio of renormalized $B$-parameters evaluated in QCD is expected to approach unity as $1/\bar\mu_h \to 0$,
This follows from standard HQET arguments, which also predict the leading
deviations for asymptotically small $1/\bar\mu_h$-values
to be of the order $1/{\rm log}(\bar\mu_h/\Lambda_{QCD})$.
Such corrections to the power scaling in $1/\bar\mu_h$
are expected to be tiny in the $\bar\mu_h$-range of our data and can be
estimated in PT by matching HQET to QCD. A possible way of doing so is to
remove them from the data by considering the modified ratio
\be
\omega_{q}(\overline\mu_h,\lambda;\overline\mu_l, a) =   
\frac{B_{B_{q}}(\overline\mu_h,\overline\mu_l, a;\mu)}{B_{B_{q}}(\overline\mu_h/\lambda,\overline\mu_l, a;\mu)}
\cdot \frac{C(\overline\mu_h;\mu^*,\mu)}{C(\overline\mu_h/\lambda;\mu^*,\mu)}\,,
\ee
where the C-factors ratio contains the info on the
$1/{\rm log}(\bar\mu_h)$-corrections at a fixed order in RG-improved PT. We
consider here HQET-to-QCD matching only at tree level and LL order in PT
(thereby avoiding the complications of operator mixing in HQET~\cite{Becirevic:2004ny}), and
confirm the impact of $1/{\rm log}(\bar\mu_h)$-corrections on the final results
to be at the level of one standard deviation~\cite{Carrasco:2012la}. 
In the left plot of fig.~\ref{fig:bag} we show the chiral-continuum limit for the ratio $\omega_s$
at the heaviest quark mass. Both the chiral and continuum extrapolation is
well under control. 
See also ref.~\cite{Carrasco:2012la}
for additional details.
In the right plot we show the heavy-quark mass dependence and the formula 
$\omega_s(\overline\mu_h) = 1 + c_1(\lambda)/\overline\mu_h + c_2(\lambda)/\overline\mu_h^2$ describes the numerical results very well.
Applying the ratio method one finally gets
\be
B_{B_s}^{\overline MS}(\overline{\mu}_b) = 0.90(5)\,,
\ee
where the error is the sum in quadrature of statistical and systematic uncertainties.
To compute $B_{B_d}$, as we have done for the decay constants, we define a ratio for $B_{B_s}/B_{B_d}$.
Also for this ratio the chiral-continuum extrapolation and the heavy-quark mass dependence is very smooth
and well under control, leading to the preliminar estimate of
\be
\frac{B_{B_s}}{B_{B_d}} = 1.03(2)\,, \qquad \qquad B_{B_d}^{\overline MS}(\overline{\mu}_b) = 0.87(5)\,.
\ee
Given this result we can also give a preliminary estimate for the parameter 
\be
\xi = \frac{f_{B_s}\sqrt{B_{B_s}}}{f_{B}\sqrt{B_{B}}}=1.21(6)\,,
\ee
where the error is the sum in quadrature of all uncertainties.
\vspace{-0.6cm}
\section*{Acknowledgements}
\label{sec:ack}
\vspace{-0.4cm}
CPU time was provided to us by the Italian SuperComputing Resource Allocation (ISCRA) 
under the class A project HP10A7IBG7 "A New Approach to B-Physics on Current Lattices" and the 
class C project HP10CJTSNF "Lattice QCD Study of B-Physics" at the CINECA supercomputing service.
We also acknowledge computer time made available to us by HLRN in Berlin.

\bibliographystyle{h-elsevier}
\bibliography{bphys}

\end{document}